\documentclass[aps,amsmath,amssymb,showpacs,unsortedaddress,longbibliography]{revtex4-1}
\usepackage{bm,amsmath,amsfonts,amssymb,graphicx,epstopdf}
\newcommand{\ud}{\mathrm{d}}
\newcommand{\xt}{\bm{x},t}
\newcommand{\xvt}{\bm{x},\bm{v},t}
\newcommand{\xyt}{\bm{x},\bm{y},t}
\newcommand{\vxt}{\vec{x},t}
\newcommand{\nmo}{\vec{n}_{\vec{m}},n_{\omega}}

\begin{document}

 \title{Eddy~diffusivity of~quasi-neutrally-buoyant~inertial~particles}

 \author{Marco \surname{Martins Afonso}}
 \affiliation{Centro de Matem\'atica da Universidade do Porto, Rua do Campo Alegre 687, 4169-007 Porto, Portugal}
 \email{marcomartinsafonso@hotmail.it}
 \author{Paolo \surname{Muratore-Ginanneschi}}
 \affiliation{Department of Mathematics and Statistics, University of Helsinki, 
Gustaf Haellstroemin katu 2b, Helsinki, Finland}
 \author{S\'{\i}lvio M.\ A.\ \surname{Gama}}
 \affiliation{Centro de Matem\'atica da Universidade do Porto, Rua do Campo Alegre 687, 4169-007 Porto, Portugal}
 \author{Andrea \surname{Mazzino}}
 \affiliation{DICCA, University of Genova, Via Montallegro 1, 16145 Genova, Italy, and INFN \& CINFAI, 
Genova Section, via Dodecaneso 33, 16146 Genova, Italy}

 \date{\today}

 \begin{abstract}
  We investigate the large-scale transport properties of quasi-neutrally-buoyant inertial particles carried by incompressible zero-mean periodic or steady
  ergodic flows. We show how to compute large-scale indicators such as the inertial-particle terminal velocity and eddy diffusivity from first principles
  in a perturbative expansion around the limit of added-mass factor close to unity. Physically, this limit corresponds to the case where the mass density
  of the particles is constant and close in value to the mass density of the fluid which is also constant. Our approach differs from the usual over-damped
  expansion inasmuch we do not assume a separation of time scales between thermalization and small-scale convection effects. For general incompressible
  flows, we derive closed-form cell equations for the auxiliary quantities determining the terminal velocity and effective diffusivity. In the special case
  of parallel flows these equations admit explicit analytic solution. We use parallel flows to show that our approach enables to shed light onto the
  behavior of terminal velocity and effective diffusivity for Stokes numbers of the order of unity. 
 \end{abstract}
 
 \maketitle

\section{Introduction}

The dynamics of inertial particles in flowing fluids (laminar and turbulent)
is an interdisciplinary research field with important consequences on climate dynamics and
hydrological cycles \cite{LF05,S07}, mainly in connection to:
global climate changes originated by PM-induced cloud formation \cite{CMT08},
and the intriguing issue related to the evidence of anomalous large fluctuations
in the residence times of mineral dust observed in different experiments carried out in the
atmosphere \cite{DCMTCGBMSSFMF16}; environmental sciences \cite{SEKH01}, in relation to pollution and deterioration of
visibility; epidemiology \cite{DPXSWFFS93}, in connection to adverse health effects in
humans; and, finally, classical fluid dynamics, e.g., to understand how a flow field
influences the particle concentration \cite{EF94,BFF01,WM03,B05,BCH07,SBMAMOP13}.

Our main aim here is to focus our attention on the large-scale transport regime
close to the limit of neutrally-buoyant particles.
This limit is relevant in a variety of situations, in way of example related to
the dispersion of particulate organic carbon in aquatic environment. By large-scale transport we mean the particle
transport dynamics observed at times large compared to those characteristic of the given velocity field.
In this limit, arguments based on the central-limit theorem suggest a diffusive regime
for particle transport characterized by effective (enhanced) diffusion coefficients \cite{K87,F95,AV95,L08}, the so-called eddy
diffusivities. These coefficients incorporate all the dynamical effects played by the
velocity field on the particle transport.
Although the diffusive scenario with eddy diffusivities is the typical one, for inertialess particles there
exist cases where superdiffusion is already observed for simple incompressible laminar
flows \cite{CMMV99} and synthetic flows \cite{ACMV00}.
Effective diffusion equations are also used in biophysics to model the substrate uptake by microorganisms in turbulently-mixed bioreactors
\cite{LMAFMS12}.

For inertialess particles, in the presence of scale separation, eddy diffusivities have been calculated by \cite{BCVV95},
exploiting a multiple-scale expansion in the scale-separation parameter.
The requirement of scale separation was relaxed by \cite{M97,MMV05,CMMV06,BMM17}, and approximate explicit
expressions for the eddy diffusivities were determined.
Conversely, for realistic flows, active on all space–-time scales, eddy diffusivities are generally
dependent on all flow characteristics and no general expression for them is known \cite{K87}.

For particles with inertia \cite{G83,MR83,BFF01,WM03,CBBBCLMT06,VCVLMPT08}, the fluid velocity does not coincide with the
particle velocity, a fact that makes the study of the large-scale transport
even more difficult than in the inertialess case. The phase space is indeed larger (it involves
both particle position and velocity) than in the inertialess case. In plain words,
the standard Fokker--Planck equation for inertialess particle concentration involving only space variables
is replaced by a Kramers equation for the case of inertial particles where
both space and velocity coordinates are involved.
Closed equations for the so-called marginal density (i.e., the physical-space particle density in which
the velocity coordinates are averaged out) have been
obtained by \cite{PS05,MAMM12} via an
over-damped expansion where a separation of time scales between thermalization and small-scale
convection effects was assumed.

Our principal objective is to propose a new type of expansion which allows one to focus on regimes arbitrarily far from the
over-damped regime to determine terminal velocity and effective diffusivity.
The expansion parameter is the departure from the limit of neutrally-buoyant particles,
in term of which small-scale co-velocity and position dynamics will be resolved on the same footing.
As we will see in detail, the main advantage of our expansion is that small-scale degrees of freedom
can be treated by means of a regular perturbation theory rather than by a secondary multi-scale expansion
as it happens for the usual over-damped expansion \cite{MAG17}.

The paper is organized as follows. In section \ref{sec:eq} we introduce the model
of inertial-particle dynamics we set out to study.
In section \ref{sec:ms} we briefly summarize the results of multi-scale
analysis of \cite{MAMM12}. We rely on these results as starting point for our analysis.
Sections \ref{sec:pt}, \ref{sec:ss}, \ref{sec:ed} report the core contributions of the present
paper. Namely, in section \ref{sec:pt} we describe the general setup for the perturbative
solution of the inertial-particle small-scale dynamics in powers of $\alpha^{1/2}$, where $\alpha$
is the deviation from added-mass ratio equal to unity. In section \ref{sec:ss} we use the expansion
to determine the equations governing the asymptotic expression of the small-scale-dynamics
asymptotic state. We use these results to determine the effective large-scale drift, the
terminal velocity, felt by the inertial particles. In section \ref{sec:ed} we show how to determine the
eddy diffusivity in an expansion in powers of $\alpha$.
In section \ref{sec:pf} we apply the general results of the previous sections to the analytically solvable
case of transport by a parallel flow. This is a useful test-bed for our methods since it offers the possibility
to contrast some of our perturbative predictions with the exact results of \cite{BMM17}.
This is what we do in section \ref{sec:comp}.
We report some further calculation details in the appendices following the conclusions.

\section{The model and definition of the large-scale problem} 
\label{sec:eq}
 
We consider a very dilute suspension of point-like inertial particles in dimension $d$, 
subject to the gravitational acceleration $\bm{g}$ and to Brownian diffusion.
The particles are carried by an incompressible velocity field $\bm{u}$ steady or periodic 
in time (with period $\mathcal{T}$), and periodic in space with unit cell $\mathbb{B}$ 
of linear size $\ell$.
For proof-of-concept purposes, we suppose that the spatial average of the velocity 
field vanishes over $\mathbb{B}$:
 \begin{equation} \label{u}
  \!\int_{\mathbb{B}}\!\ud\bm{x}\,\bm{u}(\xt)=\bm{0}\;.
 \end{equation} 
We refer the reader interested in the role played by
mean currents on large-scale transport indicators to
e.g.\ \cite{HM94,MV97,MMV05,CMMV06,FN10,MAMG16}. 

We assume that the incompressible vector field $\bm{u}$ is also ergodic. 
We will make more precise in what follows the meaning and  
consequences of this assumption.
 
We take as model for the dynamics of an individual inertial particle 
the stochastic differential equations with additive noise \cite{MR83,G83}:
 \begin{equation} \label{basic}
  \left\{\begin{array}{rcl}
   \dot{\bm{\mathcal{X}}}(t)&=&\bm{\mathcal{V}}(t)+\beta\bm{u}(\bm{\mathcal{X}}(t),t)+\sqrt{2D}\bm{\mu}(t)\;,\\
   \dot{\bm{\mathcal{V}}}(t)&=&\displaystyle-\frac{\bm{\mathcal{V}}(t)-(1-\beta)\bm{u}(\bm{\mathcal{X}}(t),t)}{\tau}
+(1-\beta)\bm{g}+\frac{\sqrt{2\kappa}}{\tau}\bm{\nu}(t)\;.
  \end{array}\right.
 \end{equation}
$\bm{\mathcal{X}}(t)$ denotes the particle position and $\bm{\mathcal{V}}(t)$ its ``co-velocity''.

In writing (\ref{basic}) we neglect any possible interaction with
other particles or with physical boundaries. The vectors
$\bm{\mu}(t)$ and $\bm{\nu}(t)$ denote independent white noises with
Brownian diffusivity constants $D$ and $\kappa$ \cite{R88}. The reason
for considering a non-vanishing Brownian force acting on the
position process is twofold. First, it naturally occurs 
in the derivation of Langevin equations for arbitrarily non-linear
systems interacting with Gaussian heat baths \cite{Z73,CK98}.  
Second, it yields a viscous regularization
of the large-scale transport equations for the particle position
process in parametric regions where homogenization analysis of
(\ref{basic}) may become ill-defined.  

The non-dimensional quantity $\beta$ in (\ref{basic}) is the ``added-mass factor''. $\beta$  
is defined as the ratio $\beta\equiv3\rho_{\mathrm{f}}/(\rho_{\mathrm{f}}+2\rho_{\mathrm{p}})$,
built from the constant fluid ($\rho_{\mathrm{f}}$) and particle
($\rho_{\mathrm{p}}$) mass densities. The role of $\beta$ is to model
the feedback of an inertial particle on the carrier flow. The feedback effectively
increases the intrinsic inertia and induces a macroscopic
discrepancy between the particle velocity $\dot{\bm{\mathcal{X}}}(t)$
and co-velocity $\bm{\mathcal{V}}(t)$. Physically-meaningful values of
$\beta$ range from $\beta=0$, for very heavy particles such as
aerosols or droplets in a gas, to $\beta=3$, for very light particles such
as bubbles in a liquid.   

Finally, in (\ref{basic}) the Stokes time $\tau$ expresses the typical
response delay of particles to flow variations. The Stokes time is defined as
$\tau\equiv\mathcal{R}^2/(3\,\eta\,\beta)$ for spherical inertial particles
of radius $\mathcal{R}$ immersed in a fluid with kinematic viscosity $\eta$.  

We suppose that the phase space of the process $(\bm{\mathcal{X}},\bm{\mathcal{V}})$ 
has the form of the Cartesian product $\check{\mathbb{B}}\times\mathbb{R}^{d}$, with 
$\mathbb{B}\subset\check{\mathbb{B}}\subseteq\mathbb{R}^{d}$. 
In other words, the co-velocity process can take unbounded values, whilst 
the position process is defined on a subset $\check{\mathbb{B}}$ of the $d$-dimensional Euclidean space. 
We are interested in situations where $\check{\mathbb{B}}$ either coincides with $\mathbb{R}^{d}$ or is a subset 
of $\mathbb{R}^{d}$ with linear size much larger than $\ell$, the linear size of the periodicity box of the 
velocity field $\bm{u}$.
We denote by $(\bm{x},\bm{v})$ the coordinates of a generic point in the process phase space.  
The particle density $\rho(\xvt):\check{\mathbb{B}}\times\mathbb{R}^{d}\times\mathbb{R}\mapsto\mathbb{R}_{+}$ 
evolves according to a Fokker--Planck equation \cite{C43,G85,R89,V07}, 
\begin{eqnarray}
\label{gfpkfk}
\mathcal{L}_{\xvt}\rho=0\;,
\end{eqnarray}
with, once we take into account that $\bm{u}$ is divergenceless,
\begin{equation} \label{kramers}
\mathcal{L}_{\xvt}=\partial_t+[\bm{v}+\beta\,\bm{u}(\xt)]\cdot\bm{\partial}_{\bm{x}}+\bm{\partial}_{\bm{v}}\cdot\left[\frac{(1-\beta)\,\bm{u}(\xt)-\bm{v}}{\tau}+(1-\beta)\bm{g}\right]-D\partial^2_{\bm{x}}-\frac{\kappa}{\tau^2}\partial^2_{\bm{v}}\;.
\end{equation}
Here and below, we adopt the convention to write in subscript only the variables on which $\mathcal{L}$ 
acts as a differential operation. 
We look for solutions of (\ref{gfpkfk}) when the largest length scale $L$ in the problem 
is either the linear size of $\check{\mathbb{B}}$ or, in the case $\check{\mathbb{B}}=\mathbb{R}^{d}$, 
the typical spatial decay scale of the initial conditions. In units of $L$, the microscopic scales of the problem are $\ell$ and 
the length scales set by the diffusivity constants and the Stokes time: $\ell_{\kappa}=\sqrt{\kappa\,\tau}$ and $\ell_{D}=\sqrt{D\,\tau}$: 
$\ell_{\kappa},\ell_{D},\ell\,\ll L$. We expect that in physically relevant situations $\ell\geq\ell_{\kappa},\ell_{D}$.
Thus, the natural quantifier of the separation between microscopic and macroscopic scales is 
the non-dimensional ``scale separation parameter''
\begin{eqnarray}
\varepsilon=\frac{\ell}{L}\ll1\;.
\nonumber
\end{eqnarray}
Our goal is to extricate the large-scale asymptotic properties, i.e.\ to inquire the
dependence of solutions upon rescaled variables $\bm{X}\equiv\varepsilon\,\bm{x}$, $T^{\ddag}=\varepsilon\,t$ 
and $T=\varepsilon^2\,t$, whilst averaging out any other functional dependence. 
We pursue our goal under the working hypothesis that the carrier vector field $\bm{u}$ 
is ergodic. By this we mean that the fundamental solution 
of (\ref{gfpkfk}) on the space of integrable functions over the $d$-dimensional torus $\mathbb{T}^{d}(\mathbb{B})$ 
specifies a contractive semigroup. Furthermore, any particular solution $\rho$ in this space
tends towards a unique steady-state distribution $p$ satisfying
\begin{eqnarray}
\label{ss}
p(\bm{x},\bm{v},t+\mathcal{T})=p(\xvt)\;.
\end{eqnarray}
This identity trivially holds for the equilibrium state of a steady flow. If we posit that
the semigroup generated by a genuinely periodic carrier field $\bm{u}$ admits a Floquet decomposition, 
then (\ref{ss}) corresponds to the eigenstate associated to unit eigenvalue of the monodromy.
In both cases, we hypothesize that the operator $\mathcal{L}^{\dag}$, adjoint to $\mathcal{L}$ 
with respect to the Lebesgue measure, admits only constants as solution of $\mathcal{L}^{\dag}f=0$
in $\mathbb{T}^{d}(\mathbb{B})\times\mathbb{R}^{d}\times[0,\mathcal{T}]$. 
The explicit uniqueness hypothesis is necessary, as the incompressibility of $\bm{u}$ does not restrict
a priori the kernel of $\mathcal{L}$ to functions constant in space --- see e.g.\ the discussion in 
\S~14.5.2 of \cite{PS07}. Given (\ref{ss}), we inquire perturbations in $\check{\mathbb{B}}\times\mathbb{R}^{d}$ of the microscopic equilibrium 
state coupled to the dynamics by the scale separation parameter $\varepsilon$.
The technical tool to perform this analysis is a multi-scale perturbation theory 
\cite{BO78,BLP78,PS07} in powers of $\varepsilon$.

\section{Multi-scale analysis}
\label{sec:ms}

The multi-scale expansion is very much along the lines of \cite{MAMM12}. 
We briefly review here the main points. 
We commence by recalling that any physical indicator may depend upon dimensional quantities
only through non-dimensional ratios. This fact allows us to infer 
that the solution of (\ref{gfpkfk}) in $\check{\mathbb{B}}\times\mathbb{R}^{d}$ can formally be written in the form
\begin{eqnarray}
\rho(\xvt)=\rho(\bm{x}_{0},\bm{X},\bm{v},t_{0},T^{\ddag},T)
\label{ms:ansatz}
\end{eqnarray}
We denote by $\bm{x}_{0}$ and $t_{0}$ any dependence of $\rho$ upon $\bm{x}$ and $t$ which does not 
appear in the form of non-dimensional ratios involving $L$. 
The idea underlying multi-scale perturbation theory is to treat 
$(\bm{X},T^{\ddag},T)$ as independent variables from $(\bm{x}_{0},t_{0})$.
Inserting (\ref{ms:ansatz}) in (\ref{kramers}) yields
\begin{eqnarray}
\mathcal{L}_{\xvt}=\mathcal{L}_{\bm{x}_{0},\bm{v},t_{0}}+\varepsilon\,\mathcal{L}^{\prime}_{\bm{x}_{0},\bm{X},T^{\ddag}}
+\varepsilon^{2}\,\mathcal{L}^{\prime\prime}_{\bm{X},T}\;,
\nonumber
\end{eqnarray}
where now
\begin{subequations}
\begin{eqnarray*}
&&\mathcal{L}^{\prime}_{\bm{x}_{0},\bm{X},T^{\ddag}}=
\partial_{T^{\ddag}}
+\Big{(}\bm{v}+\bm{u}(\bm{x}_{0},t)\Big{)}\cdot\partial_{\bm{X}}
-2\,D\,\partial_{\bm{x}_{0}}\cdot\partial_{\bm{X}}\;,
\\
&&\mathcal{L}^{\prime\prime}_{\bm{X},T}=\partial_{T}-D\,\partial_{\bm{X}}^{2}\;.
\end{eqnarray*}
\end{subequations}
The foliation allows us to look for a perturbative solution
of the form
\begin{eqnarray}
\rho(\xvt)=p(\bm{x}_{0},\bm{v},t_{0})\,P(\bm{X}-\bm{w}\,T^{\ddag},T)
+\varepsilon\,\bm{q}(\bm{x}_{0},\bm{v},t_{0})\cdot\partial_{\bm{X}}P(\bm{X}-\bm{w}\,T^{\ddag},T)
+O(\varepsilon^{2})\;.
\label{ms:sol}
\end{eqnarray}
The useful feature of this expression is to represent $\rho$ as a sum of products of functions depending 
exclusively upon small-scale variables $\bm{x}_{0},\bm{v},t_{0}$ times functions of the large-scale variables $\bm{X}$, $T^{\ddag}$, $T$ alone. 
In particular, the function $P$ encodes the large-scale asymptotics of the full solution we are after.
As usual in homogenization theory \cite{BCVV95,M97,BML16,PS05}, we determine
$P$ by canceling secular terms from the expansion up to order $O(\varepsilon^{2})$.
The upshot \cite{MAMM12} is that $P$ satisfies the diffusion equation
\begin{eqnarray}
\label{pk}
\partial_{T}P-K_{ij}\partial_{X_{i}}\partial_{X_{j}}P=0\;,
\end{eqnarray}
whilst
 \begin{equation} 
\label{dabliu}
\bm{w}
=\int_0^{\mathcal{T}}\frac{\ud t}{\mathcal{T}}\int_{\mathbb{B}}\ud\bm{x}\int_{\mathbb{R}^d}\ud\bm{v}\,
\Big{(}\bm{v}+\beta\,\bm{u}(\xt)\Big{)}\,p(\xvt)
 \end{equation}
describes the uniform effective drift acting on the particle at large scales. 
We refer to $\bm{w}$ as the terminal velocity \cite{MC86,WM93}. The effective diffusion tensor in (\ref{pk}) is
 \begin{eqnarray} 
\label{cappa}
  K_{ij}=D\,\delta_{ij}-\frac{1}{2}\sum_{\sigma}\int_0^{\mathcal{T}}\frac{\ud t}{\mathcal{T}}\int_{\mathbb{B}}\ud\bm{x}
\int_{\mathbb{R}^d}\ud\bm{v}\,\Big{(}v_{\sigma(i)}+\beta u_{\sigma(i)}(\xt)-w_{\sigma(i)}\Big{)}q_{\sigma(j)}(\xvt)\;,
 \end{eqnarray}
where the sum over $\sigma$ ranges over the permutations of (i.e., symmetrizes) the indices $i,j$.
Finally, the auxiliary vector field $\bm{q}(\xvt)$ in (\ref{ms:sol}), (\ref{cappa}) is specified
by the solution of
 \begin{equation} \label{lq}
  \mathcal{L}_{\xvt}\,\bm{q}(\xvt)
=-[\bm{v}+\beta\bm{u}(\xt)-\bm{w}]\,p(\xvt)\;,
 \end{equation}  
complemented by the solvability condition
\begin{eqnarray}
\int_0^{\mathcal{T}}\frac{\ud t}{\mathcal{T}}\int_{\mathbb{B}}\ud\bm{x}
\int_{\mathbb{R}^d}\ud\bm{v}\,\bm{q}(\xvt)=0
\label{ms:solvability}
\end{eqnarray}
in the space $\mathbb{L}^{2}(\mathbb{T}^{d+1}\times\mathbb{R}^{d})$ of functions square integrable with respect 
to the Lebesgue measure $\ud t\,\ud\bm{x}\,\ud\bm{v}$, 
periodic for $(\xt)\in\mathbb{B}\times[0,\mathcal{T}]$,
i.e.\ defined on the torus $\mathbb{T}^{d+1}\equiv\mathbb{T}^{d+1}(\mathbb{B}\times[0,\mathcal{T}])$, and normalizable 
for $\bm{v}\in\mathbb{R}^{d}$. 
To neaten the notation we drop in (\ref{dabliu}) and in what follows the subscript ``$0$'' 
for variables defined on the $d$-dimensional torus $\mathbb{T}^{d}(\mathbb{B})$.

We refer to \cite{BLP78,PS07} for further details on functional-analysis aspects of the solution space.
We also refer to \cite{MAMM12} for the derivation of (\ref{pk}) and the proof that $K_{ij}$ are indeed 
the components of a positive definite tensor $\mathsf{K}$. 

In summary, the explicit evaluation of terminal velocity and effective diffusivity 
requires the knowledge of the steady-state probability density and the solution 
of the vector equations (\ref{lq}), (\ref{ms:solvability}). The following observation is helpful 
in reference to the evaluation of the terminal velocity. In the steady state the identity
\begin{eqnarray}
0&=&\int_{0}^{\mathcal{T}}\ud t\,\partial_{t}\int_{\mathbb{B}}\ud\bm{x}\int_{\mathbb{R}^d}\ud\bm{v}\,
\bm{v}\,p(\xvt)
\nonumber\\
&=&-\int_{0}^{\mathcal{T}}\ud t\int_{\mathbb{B}}\ud\bm{x}\int_{\mathbb{R}^d}\ud\bm{v}\,
\left[-\frac{(1-\beta)\,\bm{u}(\xt)-\bm{v}}{\tau}-(1-\beta)\bm{g}\right]
\,p(\xvt)
\nonumber
\end{eqnarray}
holds true for any sufficiently regular probability preserving boundary conditions. 
We avail us of this identity to couch the steady-state terminal velocity into the 
form
\begin{eqnarray*}
\bm{w}=(1-\beta)\,\bm{g}\,\tau+\bm{W}\;,
\end{eqnarray*}
where the discrepancy with respect to the bare value in still fluids is
\begin{eqnarray}
\bm{W}\equiv\int_0^{\mathcal{T}}\frac{\ud t}{\mathcal{T}}\int_{\mathbb{B}}\ud\bm{x}\int_{\mathbb{R}^d}\ud\bm{v}\,
\bm{u}(\xt)\,p(\xvt)\;.
\label{ms:rterminal}
\end{eqnarray}
Thus, $\bm{w}$ might be non-vanishing in the absence of
gravity even for an advecting flow with zero spatial average.

\section{Perturbative expansion of the small-scale dynamics}
\label{sec:pt}

Our goal is to set up a perturbative scheme to solve (\ref{lq}) on $\mathbb{L}^{2}(\mathbb{T}^{d+1}\times\mathbb{R}^{d})$.
To do this we consider particles whose mass density differs only
slightly (either in excess or defect) from the fluid mass density
\cite{BCPP00,MFS07,SH08}. The reason is that for $\beta=1$ the steady state (\ref{ss})
reduces to the equilibrium state  
\begin{eqnarray}
\label{pt:ss}
p^{(0)}(\xvt)\equiv\frac{1}{\ell^{d}}
\left(\frac{\tau}{2\,\pi\,\kappa}\right)^{d/2}\exp\left(-\frac{\tau\,\|\bm{v}\|^{2}}{2\,\kappa}\right)\;.
\end{eqnarray}
Also, the case $\beta=1$ is in close resemblance with the situation described in \cite{CC99} for tracers.
For $\beta\simeq1$, we introduce the expansion parameter $\alpha\equiv|1-\beta|\ll1$.  
It is also expedient to define $J\equiv\mathrm{sgn}(1-\beta)$ and to suppose that the ratio
\begin{eqnarray}
k\equiv\frac{\kappa}{\alpha}
\nonumber
\end{eqnarray}
is independent of $\alpha$. In analogy to \cite{MA08,MAM11,MAMM12}, we introduce the 
change of co-velocity coordinates 
\begin{eqnarray}
\bm{v}=\alpha^{1/2}\,\bm{y}+J\,\alpha\,\bm{g}\,\tau\;,
\nonumber
\end{eqnarray}
and we correspondingly define the Gaussian measure with density
\begin{eqnarray}
G(\bm{y})=\left(\frac{\tau}{2\,\pi\,k}\right)^{d/2}\exp\left(-\frac{\tau\,\|\bm{y}\|^{2}}{2\,k}\right)\;.
\label{pt:Gaussian}
\end{eqnarray}
We use $G$ to perform a similarity transformation on the Fokker--Planck
operator $\mathcal{L}$. The result is 
 \begin{equation}
G^{-1}(\bm{y})\mathcal{L}_{\xvt}G(\bm{y})
=\mathcal{L}^{(0)}_{\xyt}
+\alpha^{1/2}\,\mathcal{L}^{(1)}_{\bm{x},\bm{y}}
+\alpha\,\mathcal{L}^{(2)}_{\bm{x},\bm{y}}
\;,
\label{pt:foliation}
\end{equation} 
with
\begin{subequations} 
\label{l012}
  \begin{eqnarray}
\mathcal{L}^{(0)}_{\xyt}&=&\partial_t+\bm{u}(\xt)\cdot\bm{\partial}_{\bm{x}}-D\,\partial^2_{\bm{x}}
+\frac{\bm{y}}{\tau}\cdot\bm{\partial}_{\bm{y}}-\frac{k}{\tau^{2}}\,\,\partial^2_{\bm{y}}\;,
\label{l012a}\\
\mathcal{L}^{(1)}_{\xyt}&=&\bm{y}\cdot\bm{\partial}_{\bm{x}}
+\frac{J}{\tau}\,\bm{u}(\xt)\cdot\left(\bm{\partial}_{\bm{y}}-\frac{\tau\,\bm{y}}{k}\right)\;, 
\label{l012b}\\
\mathcal{L}^{(2)}_{\xyt}&=&J\,[\bm{g}\,\tau-\bm{u}(\xt)]\cdot\bm{\partial}_{\bm{x}}\;.
\label{l012c}
  \end{eqnarray}
 \end{subequations}  
The decomposition (\ref{l012}) shows that for $\alpha\,\ll\,1$ we may describe the dynamics of 
perturbations of the equilibrium state (\ref{pt:ss}) in terms of the operator (\ref{l012a}) 
acting on $\mathbb{L}^{2}(\mathbb{T}^{d+1}\times\mathbb{R}^{d},G(\bm{y})
\ud\bm{x}\,\ud\bm{y}\,\ud t)$, i.e.\ the space of square-integrable functions 
with respect to the Gaussian measure specified by (\ref{pt:Gaussian}). 
The advantage of this description is that (\ref{l012a}) is of the form
\begin{eqnarray}
\mathcal{L}^{(0)}_{\xvt}=\partial_{t}+\mathcal{M}_{\bm{x}}-\tau^{-1}\mathcal{N}_{\bm{y}}
\label{pt:l0}
\end{eqnarray}
where
 \begin{equation}
 \label{m}
\mathcal{M}_{\bm{x}}\equiv\bm{u}(\xt)\cdot\bm{\partial}_{\bm{x}}-D\,\partial^2_{\bm{x}}
 \end{equation} 
is the generator of an advection--diffusion process in position
space and
 \begin{equation}
 \label{n}
\mathcal{N}_{\bm{y}}\equiv-\bm{y}\cdot\bm{\partial}_{\bm{y}}+k\,\tau^{-1}\partial^2_{\bm{y}}
 \end{equation}  
the generator of an Ornstein--Uhlenbeck process in co-velocity space.  
Hence, we can use the method of separation of constants to factorize the 
dependence between position and co-velocity coordinates in order to solve them 
in two separate steps. 

The dynamics in the co-velocity space is exactly integrable. Specifically, the 
Ornstein--Uhlenbeck operator (\ref{n}) is self-adjoint in $\mathbb{L}^{2}(\mathbb{R}^{d},G(\bm{y})\ud\bm{y})$
(i.e.\ with respect to the equilibrium measure). The spectrum of $\mathcal{N}_{\bm{y}}$
coincides with $\mathbb{N}$ so that for each $n\in\mathbb{N}$ the corresponding 
eigenvectors are Hermite polynomials $\mathsf{H}^{(n)}$ of order $n$ in $d$-dimensions (see e.g.\ \cite{G49} 
and also appendix~\ref{ap:Hermite}). 
Based on this fact, we reduce the problem of inverting (\ref{l012a}) to that 
of determining the spectral properties of the operator $\breve{\mathcal{M}}_{\xt}$ acting on square integrable
functions on the torus $\mathbb{T}^{d+1}(\mathbb{B}\times[0,\mathcal{T}])$, i.e.\ in formulas
\begin{eqnarray}
\breve{\mathcal{M}}_{\xt}f=(\partial_{t}+\mathcal{M}_{\bm{x}})f\,,\hspace{0.5cm}\forall\,f\,\in\,
\mathbb{L}^{2}(\mathbb{T}^{d+1})\;.
\label{pt:operator}
\end{eqnarray}
Heuristic considerations based on Floquet theory \cite{K82} together
with the ergodicity assumption give us qualitative information about the spectral
properties of $\breve{\mathcal{M}}_{\xt}$ and its adjoint in $\mathbb{L}^{2}(\mathbb{T}^{d+1})$. 
The property that we need is that, in consequence of the ergodicity of $\bm{u}$, 
the elements of the kernel of $\breve{\mathcal{M}}_{\xt}$ and $\breve{\mathcal{M}}^{\dag}_{\xt}$
consist only of constant functions.

Based on the above considerations we expand the steady-state measure as
\begin{eqnarray}
p(\xvt)=\sum_{\mathcal{K}=0}^{\infty}\alpha^{\mathcal{K}/2}\frac{G(\bm{y})}{\ell^{d}}\,p^{(\mathcal{K})}(\xyt)\;,
\label{pt:}
\end{eqnarray}
with $p^{(0)}(\xyt)=1$. The addends in the expansion satisfy the hierarchy of equations 
  \begin{subequations} 
\label{01K}
  \begin{eqnarray}
   \mathcal{L}^{(0)}p^{(1)}&=&\frac{J}{k}\bm{y}\cdot\bm{u}\;, \label{01Kb}\\
   \mathcal{L}^{(0)}p^{(\mathcal{K})}&=&-\mathcal{L}^{(1)}p^{(\mathcal{K}-1)}-\mathcal{L}^{(2)}p^{(\mathcal{K}-2)}\qquad(\mathcal{K}\ge2)\;, \label{01Kc}
  \end{eqnarray}
 \end{subequations}
subject to the conditions
\begin{eqnarray}
\int_{\mathbb{R}^d}\ud\bm{v}\,G(\bm{y})\,p^{(\mathcal{K})}(\xyt)=0\hspace{1.0cm}\forall\,\mathcal{K}>0\;.
\label{ms:pbc}
\end{eqnarray}
Similarly, we expand the auxiliary vector
 \begin{equation}
  \bm{q}(\xvt)=
\sum_{\mathcal{K}=0}^{\infty}\alpha^{\mathcal{K}/2}\frac{G(\bm{y})}{\ell^{d}}\,\bm{q}^{(\mathcal{K})}(\xyt)\;,
\label{pt:aux}
 \end{equation}
and find the hierarchy
 \begin{subequations} 
\label{01J}
  \begin{eqnarray}
   \mathcal{L}^{(0)}\bm{q}^{(0)}&=&-(\bm{u}-\bm{W}^{(0)})p^{(0)}\;, \label{01Ja}\\
   \mathcal{L}^{(0)}\bm{q}^{(1)}&=&-\mathcal{L}^{(1)}\bm{q}^{(0)}-(\bm{u}-\bm{W}^{(0)})\,p^{(1)}-(\bm{y}-\bm{W}^{(1)})\,p^{(0)}\;, \label{01Jb}\\
   \mathcal{L}^{(0)}\bm{q}^{(\mathcal{K})}&=&-\mathcal{L}^{(1)}\bm{q}^{(\mathcal{K}-1)}-\mathcal{L}^{(2)}\,\bm{q}^{(\mathcal{K}-2)}
    -\bm{u}\,p^{(\mathcal{K})}-\bm{y}\,p^{(\mathcal{K}-1)}+J\,\bm{u}\,p^{(\mathcal{K}-2)}\nonumber\\
   &&+\sum_{\mathcal{J}=0}^{\mathcal{K}}\bm{W}^{(\mathcal{J})}\,p^{(\mathcal{K}-\mathcal{J})}\qquad(\mathcal{K}\ge2)\;, \label{01Jc}
  \end{eqnarray}
 \end{subequations}
complemented for any $\mathcal{K}$ by equations for the expansion of the reduced terminal velocity in
(\ref{ms:rterminal}),
\begin{eqnarray}
\bm{W}^{(\mathcal{K})}=
\int_0^{\mathcal{T}}\frac{\ud t}{\mathcal{T}}\int_{\mathbb{B}}\ud\bm{x}\int_{\mathbb{R}^d}\ud\bm{v}\,
\bm{u}(\xt)\,p^{(\mathcal{K})}(\xyt)\;,
\nonumber
\end{eqnarray}
and by the set of solvability conditions 
\begin{eqnarray}
\label{pt:solvability}
\int_0^{\mathcal{T}}\ud t\int_{\mathbb{B}}\ud\bm{x}
\int_{\mathbb{R}^d}\ud\bm{v}\,G(\bm{y})\,\bm{q}^{(\mathcal{K})}(\xyt)=0\;.
\end{eqnarray}
An analysis of the hierarchies (\ref{01K}), (\ref{01J}) leads to further simplifications. 
Namely, it is expedient to decompose the $\bm{y}$ dependence of the $p^{(\mathcal{K})}$'s and $\bm{q}^{(\mathcal{K})}$'s 
into Hermite polynomials. We then see that $\mathcal{L}^{(1)}$ acts on 
the space $\mathcal{H}_{n}$ of Hermite polynomials of order $n$ as a linear combination 
of one unit raising and lowering operators,
\begin{eqnarray}
\mathcal{L}^{(1)}:\mathcal{H}_{n}\mapsto\mathcal{H}_{n-1}\oplus\mathcal{H}_{n+1}\;,
\nonumber
\end{eqnarray}
whilst $\mathcal{L}^{(2)}$ preserves $\mathcal{H}_{n}$:
\begin{eqnarray}
\mathcal{L}^{(2)}:\mathcal{H}_{n}\mapsto\mathcal{H}_{n}\;.
\nonumber
\end{eqnarray}
This means that the hierarchy (\ref{01K}) couples only spaces $\mathcal{H}_{n}$
with the same parity. We are therefore entitled to look for solutions 
of the form (Einstein's convention on index contractions):
\begin{subequations}
\label{pt:pexp}
\begin{eqnarray}
&&p^{(2\mathcal{K})}(\xyt)=\sum_{\mathcal{J}=0}^{\mathcal{K}}
H_{i_{1},\dots,i_{2\mathcal{J}}}^{(2\mathcal{J})}(\bm{y})\,P^{(2\mathcal{K}:2\mathcal{J})}_{i_{1},\dots,i_{2\mathcal{J}}}(\xt)\;,
\label{pt:pexp1}
\\
&&p^{(2\mathcal{K}+1)}(\xyt)=\sum_{\mathcal{J}=0}^{\mathcal{K}}
H_{i_{1},\dots,i_{2\mathcal{J}+1}}^{(2\mathcal{J}+1)}(\bm{y})\,P^{(2\mathcal{K}+1:2\mathcal{J}+1)}_{i_{1},\dots,i_{2\mathcal{J}+1}}(\xt)\;.
\label{pt:pexp2}
\end{eqnarray}
\end{subequations}
For the reduced terminal velocity this result means
\begin{subequations}
\label{pt:terminal}
\begin{eqnarray}
&&\bm{W}^{(2\,\mathcal{K})}=
\int_0^{\mathcal{T}}\frac{\ud t}{\mathcal{T}}\int_{\mathbb{B}}\frac{\ud\bm{x}}{\ell^{d}}\,
\bm{u}(\xt)\,P^{(2\mathcal{K}:0)}(\xt)\;,
\\
&&\bm{W}^{(2\,\mathcal{K}+1)}=\bm{0}\;,
\end{eqnarray}
\end{subequations}
which, on its turn, implies that also the hierarchy for the auxiliary vector field couples 
only spaces $\mathcal{H}_{n}$ with the same parity. 
As the right-hand sides of (\ref{01Ja}) and (\ref{01Jb}) are respectively 
independent and linear in the co-velocity variables, we conclude that also the $\bm{q}^{(\mathcal{K})}$'s 
admit an expansion similar to (\ref{pt:pexp}):
\begin{subequations}
\label{pt:qexp}
\begin{eqnarray}
&&q^{(2\mathcal{K})}_{i}(\xyt)=\sum_{\mathcal{J}=0}^{\mathcal{K}}
H_{i_{1},\dots,i_{2\mathcal{J}}}^{(2\mathcal{J})}(\bm{y})\,Q^{(2\mathcal{K}:2\mathcal{J})}_{i,i_{1},\dots,i_{2\mathcal{J}}}(\xt)\;,
\label{pt:qexp1}
\\
&&q^{(2\mathcal{K}+1)}_{i}(\xyt)=\sum_{\mathcal{J}=0}^{\mathcal{K}}
H_{i_{1},\dots,i_{2\mathcal{J}+1}}^{(2\mathcal{J}+1)}(\bm{y})\,Q^{(2\mathcal{K}+1:2\mathcal{J}+1)}_{i,i_{1},\dots,i_{2\mathcal{J}+1}}(\xt)\;.
\label{pt:qexp2}
\end{eqnarray}
\end{subequations}
We also notice that (\ref{ms:pbc}) and (\ref{ms:solvability}) yield the conditions
\begin{eqnarray}
\int_{\mathbb{B}}\frac{\ud\bm{x}}{\ell^{d}}\,P^{(2\mathcal{K}:0)}(\xt)=\int_{\mathbb{B}}\frac{\ud\bm{x}}{\ell^{d}}\,Q^{(2\mathcal{K}:0)}_i(\xt)=0\;.
\nonumber
\end{eqnarray} 
These considerations imply that also the eddy-diffusivity tensor admits an expansion 
in powers of $\alpha$ rather than $\alpha^{1/2}$: 
\begin{equation} 
\label{pt:K}
\mathsf{K}=D\,\mathsf{I}+\sum_{\mathcal{J}=0}^{\infty}\alpha^{\mathcal{J}}\,\mathsf{K}^{(2\mathcal{J})}\;;
 \end{equation}
here,
\begin{equation}
K^{(2\mathcal{J})}_{ij}=
-\frac{1}{2} \sum_{\sigma}\int_0^{\mathcal{T}}\frac{\ud t}{\mathcal{T}}\int_{\mathbb{B}}\frac{\ud\bm{x}}{\ell^{d}}
 \Big{(}u_{\sigma(i)}(\xt)(Q^{(2\mathcal{J}:0)}_{\sigma(j)}(\xt)-J\,Q^{(2\mathcal{J}-2:0)}_{\sigma(j)}(\xt))+\frac{k}{\tau}\,Q^{(2\mathcal{J}-1:1)}_{\sigma(1)\sigma(j)}(\xt)\Big{)}
\label{pt:kij}
\end{equation}
with the convention $Q^{(k:0)}_{\dots}=0$ for any $k\,<\,0$.

We are thus left with the task of proving that the equations for the $P$'s and $Q$'s 
are generically well posed.

\section{Steady state and terminal velocity} 
\label{sec:ss}

 $P^{(0:0)}=1$ and the hypothesis (\ref{u}) immediately imply
\begin{eqnarray}
\bm{W}^{(0)}=\bm{0}\;.
\nonumber
\end{eqnarray}
The leading-order correction to the steady state satisfies 
\begin{eqnarray}
  \Big{(}\breve{\mathcal{M}}_{\xt}+\tau^{-1}\Big{)}P^{(1:1)}_{i}=
\frac{J}{k}u_{i}\;,
  \label{ss:P1}
\end{eqnarray}
whilst the second order is governed by
\begin{subequations}
\label{ss:P2}
\begin{eqnarray}
&&\breve{\mathcal{M}}_{\xt}\,P^{(2:0)}=
-\frac{k}{\tau} \partial_{x_{i}} P^{(1:1)}_{i}\;,
\label{ss:P20}\\
&&\Big{(}\breve{\mathcal{M}}_{\xt}+\frac{2}{\tau}\,\Big{)}P^{(2:2)}_{i j}=
-\frac{1}{2}\sum_{\sigma}\left(\partial_{x_{\sigma(i)}}P^{(1:1)}_{\sigma(j) }
-\frac{J}{k}u_{\sigma(i)}\,P^{(1:1)}_{\sigma(j) }\right)\;.
\label{ss:P22}
\end{eqnarray}
\end{subequations}
The left-hand sides of (\ref{ss:P1}), (\ref{ss:P2}) exhibit the general feature of the hierarchy. 
In order to determine the multi-tensor $P^{(\mathcal{K}:\mathcal{J})}_{i_{1},\dots,i_{l}}$ we need to invert the operator
$\breve{\mathcal{M}}_{\xt}+\mathcal{J}\,\tau^{-1}$
in $\mathbb{L}^{2}(\mathbb{T}^{d+1})$. For $\mathcal{J}=0$ we need to check that non-homogeneous terms on the right-hand side of (\ref{pt:pexp}) 
be orthogonal to the kernel of $\breve{\mathcal{M}}^{\dag}_{\xt}$. More explicitly, this means that the non-homogeneous terms must have 
zero average with respect to the uniform measure in $\mathbb{B}$.
For $\mathcal{J}\,>\,0$ we notice instead that the term $\mathcal{J}\,\tau^{-1}$ simply results in a positive 
shift of the real part of the spectrum of $\breve{\mathcal{M}}_{\xt}$. We expect therefore that
for $\mathcal{J}\,>\,0$ we generically need to invert operators with empty kernel. Based on this inference we conclude 
that the expansion (\ref{pt:aux}) in powers of $\alpha^{1/2}$ does not bring about secular terms.

As a further illustration of the expansion, we list also the equations determining the
$O(\alpha^{2})$ correction to the terminal velocity. From the $O(\alpha^{3/2})$ we need
\begin{equation}
\left(\breve{\mathcal{M}}_{\xt}+\frac{1}{\tau}\right)P^{(3:1)}_{i}=
-\left(\partial_{x_{i}}-\frac{J}{k} u_{i}\right)\,P^{(2:0)}
-J\,(\bm{g}\,\tau-\bm{u})\cdot\partial_{\bm{x}}\,P^{(1:1)}_{i}-\frac{2\,k}{\tau}
\partial_{x_{i}}P^{(2:2)}_{ij}\;,
\label{ss:P31}
\end{equation}
as $P^{(3:1)}_{i}$ enters the non-homogeneous term in
\begin{equation*}
\breve{\mathcal{M}}_{\xt}P^{(4:0)}
=-J\,(\bm{g}\,\tau-\bm{u})\cdot\partial_{\bm{x}}P^{(2:0)}
-\frac{k}{\tau}\partial_{\bm{x}_{i}}\,P^{(3:1)}_{i}\;.
\end{equation*}

\section{Eddy diffusivity} 
\label{sec:ed}

The results of the previous section imply that (\ref{01Ja}) reduces to
 \begin{equation}
  \breve{\mathcal{M}}_{\xt}\bm{Q}^{(0:0)}(\xt)=-\bm{u}\;.
\label{ed:zero}
 \end{equation}
By the hypothesis (\ref{u}) the carrier field $\bm{u}$ is orthogonal to the 
kernel of $\breve{\mathcal{M}}^{\dag}_{\xt}$, and so is $\bm{Q}^{(0:0)}$ by (\ref{pt:solvability}). Hence the problem of
expressing $\bm{Q}^{(0:0)}$ in terms of $\bm{u}$ is well-posed.
Conversely, the expression of $\bm{u}$ in terms 
of $\bm{Q}^{(0:0)}$ allows us to prove that the leading order of the expansion 
in powers of $\alpha$ of the eddy diffusivity is positive definite. Namely, the identity 
 \begin{eqnarray} 
\label{K0}
K^{(0)}_{ij}&=&-\frac{1}{2}\sum_{\sigma}\int_0^{\mathcal{T}} \frac{\ud t}{\mathcal{T}}\int_{\mathbb{B}}\frac{\ud\bm{x}}{\ell^{d}}
u_{\sigma(i)}(\xt)\,Q^{(0:0)}_{\sigma(j)}(\xt)
\nonumber\\
&=&D\,\int_0^{\mathcal{T}}\frac{\ud t}{\mathcal{T}}\int_{\mathbb{B}}\frac{\ud\bm{x}}{\ell^{d}}\,
(\partial_{x_{l}}Q^{(0)}_{i})(\xt)\,(\partial_{x_{l}}Q^{(0)}_{j})(\xt)
 \end{eqnarray}
holds true. This is because we take advantage of the symmetry of the tensor and 
the periodicity of the integrand to prove that total derivatives give vanishing contribution to the
integral. 

Gravity appears in (\ref{l012}) only through $\mathcal{L}^{(2)}$. Thus, we need to compute at least the 
order $O(\alpha)$ in the expansion (\ref{pt:aux}) of the auxiliary vector field $\bm{q}$ in order to inquire
how $\bm{g}$ affects the eddy diffusivity.

We avail us of (\ref{ed:zero}) to solve (\ref{01Jb}). We find
\begin{eqnarray}
  \Big{(}\breve{\mathcal{M}}_{\xt}+\frac{1}{\tau}\Big{)}Q^{(1:1)}_{ij}=
-\left(\partial_{x_{j}}-\frac{J}{k}\,u_{j}\right)Q_{i}^{(0:0)}-u_{i}\,P^{(1:1)}_{j}-\delta_{ij}\;.
  \label{ed:Q1}
\end{eqnarray}
 $Q^{(1:1)}_{ij}$ specifies the non-homogeneous term in the equations for the $O(\alpha)$ corrections:
\begin{subequations}
\label{ed:Q2}
\begin{eqnarray}
\breve{\mathcal{M}}_{\xt}\,Q^{(2:0)}_{i}&=&-J(\bm{g}\,\tau-\bm{u})\cdot\partial_{\bm{x}}Q^{(0:0)}_{i}
\nonumber\\&&
-\frac{k}{\tau}\partial_{x_{j}}Q^{(1:1)}_{ij}-\frac{k}{\tau}P^{(1:1)}_{i}-u_{i}\,P^{(2:0)}+J\,u_{i}+W^{(2)}_{i}
\label{ed:Q20}
\\
\Big{(}\breve{\mathcal{M}}_{\xt}+\frac{2}{\tau}\,\Big{)}\,Q^{(2:2)}_{ijl}&=&
-\frac{1}{2}\sum_{\sigma}\left(\partial_{x_{\sigma(l)}}Q^{(1:1)}_{i\,\sigma(j)}
-\frac{J}{k}u_{\sigma(l)}\,Q^{(1:1)}_{i\,\sigma(j) }+u_{\sigma(l)}\,P^{(2:2)}_{i\,\sigma(j)}\right)-P^{(1:1)}_{i}\delta_{jl}\;.\hspace{1cm}
\label{ed:Q22}
\end{eqnarray}
\end{subequations}
We readily verify that the right-hand side of (\ref{ed:Q20}) is orthogonal to the kernel of
$\breve{\mathcal{M}}^{\dag}_{\xt}$. We then use (\ref{ed:Q1}) to eliminate $Q^{(1:1)}$ from the expression 
of the leading-order correction to the eddy diffusivity:
 \begin{equation} 
\label{K2}
  K^{(2)}_{ij}=k\,\delta_{ij}+\frac{1}{2}\sum_{\sigma}\int_0^{\mathcal{T}}\frac{\ud t}{\mathcal{T}}
\int_{\mathbb{B}}\frac{\ud\bm{x}}{\ell^{d}}\,u_{\sigma(i)}(\xt)\,\Big{(}k\,P^{(1:1)}_{j}(\xt)-Q^{(2:0)}_{\sigma(j)}(\xt)\Big{)}\;.
 \end{equation}
A further simplification occurs if we introduce $R_{i}^{(2:0)}\equiv Q_{i}^{(2:0)}-k\,P_{i}^{(1:1)}$. Using (\ref{ss:P1}) 
it is straightforward to verify that:
\begin{eqnarray}
\breve{\mathcal{M}}_{\xt}\,R^{(2:0)}_{i}=
-J(\bm{g}\,\tau-\bm{u})\cdot\partial_{\bm{x}}Q^{(0:0)}_{i}
-\frac{k}{\tau}\partial_{x_{j}}Q^{(1:1)}_{i\,j}-u_{i}\,P^{(2:0)}+W^{(2)}_{i}\;.
\label{ed:Q20tilde}
\end{eqnarray}
We arrive at:
\begin{eqnarray}
 K^{(2)}_{ij}=k\,\delta_{ij}-\frac{1}{2}\sum_{\sigma}\int_0^{\mathcal{T}}\frac{\ud t}{\mathcal{T}}
\int_{\mathbb{B}}\frac{\ud\bm{x}}{\ell^{d}}\,u_{\sigma(i)}(\xt)\,R^{(2:0)}_{\sigma(j)}(\xt)\;.
\nonumber
\end{eqnarray}
In general, we are not able to derive the explicit dependence of (\ref{K2}) upon gravity. 
Our conclusion is that the eddy diffusivity depends upon $\bm{g}$ non-trivially
through the explicit form of the solution of (\ref{ed:Q20tilde}).

Finally, we write the equations governing the $O(\alpha^{2})$ contribution to the eddy diffusivity. 
As in the case of the expansion of the steady state we need
\begin{eqnarray}
\left(\breve{\mathcal{M}}_{\xt}+\frac{1}{\tau}\right)Q^{(3:1)}_{ij}
&=&-\left(\partial_{x_{j}}-\frac{J}{k}u_{j}\right)\,Q^{(2:0)}_{i}
-J(\bm{g}\,\tau-\bm{u})\cdot\partial_{\bm{x}}\,Q^{(1:1)}_{ij}
-\frac{k}{\tau}\partial_{x_{l}}(Q^{(2:2)}_{ilj}+Q^{(2:2)}_{ijl})
\nonumber\\&&
-u_{j}\,P^{(3:1)}_{i}-\delta_{ij}\,P^{(2:0)}+(J\,u_{j}+W^{(2)}_{j})\,P^{(1:1)}_{i}
-\frac{2\,k}{\tau}\,P^{(2:2)}_{ij}
\label{ed:Q301}
\end{eqnarray}
in order to then specify all terms entering
 \begin{eqnarray}
\breve{\mathcal{M}}_{\xt}Q^{(4:0)}_i&=&-\,J\,(\bm{g}\,\tau-\bm{u})\cdot\partial_{\bm{x}}Q^{(2:0)}_i
-\frac{k}{\tau}\partial_{x_{j}}\,Q^{(3:1)}_{j i}
\nonumber\\&&
-\frac{k}{\tau}P^{(3:1)}_{i}-u_{i}\,P^{(4:0)}+(J\,u_{i}+W^{(2)}_{i})\,P^{(2:0)}+W^{(4)}_{i}\;.
   \label{ed:Q4}
 \end{eqnarray}
Lastly, we use (\ref{ed:Q301}) and (\ref{ss:P22}) to simplify the expression of the 
contribution to the eddy diffusivity. We obtain:
 \begin{eqnarray} \label{K4}
  K^{(4)}_{ij}=\frac{1}{2}\sum_{\sigma}\int_{0}^{\mathcal{T}}\frac{\ud t}{\mathcal{T}}\int_{\mathbb{B}}
\frac{\ud\bm{x}}{\ell^{d}}\,u_{\sigma(i)}(\xt)\,\Big{(}k\,P^{(3:1)}_{\sigma(j)}(\xt)
-Q^{(4:0)}_{\sigma(j)}(\xt)\Big{)}\;.
 \end{eqnarray}
Again, we can simplify this expression by introducing $R^{(4:0)}_{i}\equiv Q^{(4:0)}_{i}-k\,P^{(3:1)}_{i}$.
In virtue of (\ref{ss:P31}), we then verify that $R^{(4:0)}_{i}$ is solution of
\begin{eqnarray}
\breve{\mathcal{M}}_{\xt}R^{(4:0)}_i&=&-J\,(\bm{g}\,\tau-\bm{u})\cdot\partial_{\bm{x}}Q^{(2:0)}_i
-\frac{k}{\tau}\partial_{\bm{x}_{j}}\,Q^{(3:1)}_{i j}+\partial_{x_{i}}\,P^{(2:0)}
\nonumber\\&&
+J(\bm{g}\,\tau-\bm{u})\cdot\partial_{\bm{x}}\,P^{(1:1)}_{i}+\frac{2\,k}{\tau}
\partial_{x_{i}}P^{(2:2)}_{ij}-u_{i}\,P^{(4:0)}+W_{i}^{(2)}\,P^{(2:0)}+W_{i}^{(4)}\;.
   \label{ed:Q4tilde}
\end{eqnarray}
Higher-order terms are amenable to similar forms. 

The results of this sections are the main findings of
the present paper. In the coming sections we illustrate their application to special 
analytically-tractable cases.

\section{Applications to parallel flows} 
\label{sec:pf}

A flow is called \emph{parallel} if it points everywhere and always into the same direction, let us say $x_1$. 
In such a case, incompressibility requires the field to be independent of $x_1$:
 \begin{equation*}
  \bm{u}(\xt)=\underline{\bm{x}}_1u(\vxt)\;,
 \end{equation*}
where $\underline{\bm{x}}_1$ is the unit vector along $x_1$, and $\vec{x}\equiv(x_2,\ldots,x_d)$ is a $(d-1)$-dimensional vector in the orthogonal
hyper-plane. Several simplifications take place for this kind of flows, due to the disappearance of the advective term in 
every application of the operator $\mathcal{M}_{\bm{x}}$ (\ref{m}), as further discussed in appendix~\ref{ap:parallel}.
 
Upon introducing the $(d-1)$-spatial and temporal Fourier transform
 \begin{eqnarray*}
  \hat{u}(\nmo)=\int\frac{\ud t}{\mathcal{T}}\int\frac{\ud\vec{x}}{\ell^{d-1}}\exp[-\imath(\vec{m}\cdot\vec{x}+\omega t)]u(\vxt)
%\ \Longleftrightarrow\ u(\vxt)=\sum_{\vec{n}_{\vec{m}}\in\mathbb{Z}^{d-1}}
%\sum_{n_{\omega}\in\mathbb{Z}}\ue^{\imath(\vec{m}\cdot\vec{x}+\omega t)}\hat{u}(\nmo)
 \end{eqnarray*}
with $\vec{m}\equiv2\pi\vec{n}_{\vec{m}}/\ell$ and $\omega\equiv2\pi n_{\omega}/\mathcal{T}$, all differential equations turn into algebraic ones and
can be solved in the Fourier space.  
In particular, (\ref{K0}) becomes:
 \begin{equation} 
\label{pf:K0}
  K^{(0)}_{ij}=\delta_{i1}\delta_{j1}\sum_{\vec{n}_{\vec{m}},n_{\omega}}D\,m^2(\omega^2+D^2m^4)^{-1}|\hat{u}(\nmo)|^2\;.
 \end{equation}
We thus recover the result of \cite{BCVV95} for the eddy diffusivity of a tracer advected by a parallel flow.

Keeping in mind that $\hat{u}(\vec{0},n_{\omega})=0$ because of (\ref{u}), we can also explicitly compute the contribution 
to the effective diffusivity in (\ref{K2}):
 \begin{equation} 
\label{pf:K2}
  K^{(2)}_{i j}=k\,\delta_{i j}
+\delta_{i1}\delta_{j1}k\sum_{\vec{n}_{\vec{m}},n_{\omega}}
\frac{m^2\tau^{-1}(\omega^2\tau^{-1}+3\omega^2D\,m^2-D^2m^4\tau^{-1}-D^3m^6)}
{(\omega^2+D^2m^4)^2[\omega^2+(D\,m^2+\tau^{-1})^2]}|\hat{u}(\nmo)|^2\;.
 \end{equation} 
Gravity does not play any role at this order, due to the reality condition of the flow field 
which by parity symmetry cancels any contribution linear in $\vec{m}$.

Finally, (\ref{K4}) becomes:
 \begin{eqnarray} 
\label{pf:K4}
  K^{(4)}_{i j}&=&\delta_{i1}\delta_{j1}\sum_{\vec{n}_{\vec{m}},n_{\omega}}|\hat{u}(\nmo)|^2
\left\{-J^2\tau^2(\vec{g}\cdot\vec{m})^2\frac{D\,m^2(3\omega^2-D^2m^4)}{(\omega^2+D^2m^4)^3}\right.\\
  &&\left.+2k^2\tau^{-3}\omega^4m^4\frac{2\tau^{-2}+5\omega^2+D\,m^2\omega^{-4}\tau^{-3}C}
{(\omega^2+D^2m^4)^3[\omega^2+(D\,m^2+\tau^{-1})^2]^2[\omega^2+(D\,m^2+2\tau^{-1})^2]}\right\}\;,\nonumber
 \end{eqnarray}
 where the function $C$ is
 \begin{eqnarray*}
\label{ed:C}
  C&=&3D^6m^{12}\tau^4/2+7D^5m^{10}\tau^3-(27\omega^2\tau^2-23)D^4m^8\tau^2/2-(45\omega^2\tau^2-8)D^3m^6\tau-(15\omega^4\times\\
  &&\times\tau^4/2+55\omega^2\tau^2-2)D^2m^4-15(\omega^2\tau^2+2)D\,m^2\omega^2\tau+(15\omega^4\tau^4/2-5\omega^2\tau^2/2-6)\omega^2\;.
 \end{eqnarray*}
It is useful to summarize our findings in a more conceptual form.
Upon recalling that $\kappa=\alpha\,k$, we can couch the eddy-diffusivity tensor in the 
form
 \begin{eqnarray} \label{chei}
  \mathsf{K}&=&(\kappa+D)\,\mathsf{I}+\underline{\bm{x}}_1\otimes\underline{\bm{x}}_1\sum|\hat{u}|^2\,F_{\emptyset}(D)
\nonumber\\&&
+\underline{\bm{x}}_1\otimes\underline{\bm{x}}_1\sum|\hat{u}|^2\left[\kappa\,F_{I}(\tau,D)+(1-\beta)^2\,g^2\,\tau^2\,F_{2:1}(D)+\kappa^2\,F_{2:2}(\tau,D)\right]
+O(\alpha^{3})\;.
\end{eqnarray}   
The representation exhibits that the eddy-diffusivity tensor differs from the Brownian 
diffusivity only for the component $K_{11}$. Such component is parallel to the flow but may 
have any orientation with respect to the gravitational acceleration. 

The $F$'s functions can be reconstructed from (\ref{pf:K0}), (\ref{pf:K2}) and (\ref{pf:K4}) in the general case 
of an incompressible velocity field. In subsections~\ref{sec:constant} and \ref{sec:random}, we give fully 
explicit expressions of the $F$'s in the case of Kolmogorov flows.  
In general, we emphasize that the second-order terms $F_{2:1}$, $F_{2:2}$ are coupled to the expansion in two distinct ways. 

The $F_{2:1}$ term is proportional to both the square
of gravity and to $\alpha^2J^2=(1-\beta)^2$, meaning that the diffusion tensor depends on the size of the
deviation from unity of the added-mass factor but not on its sign. Gravity enters the eddy diffusivity 
only via this coupling, as foreseeable from the model equations (\ref{basic}).

The $F_{2:2}$ term is proportional to $\alpha^2k^2=\kappa^2$ and depends on $\tau$ in a complicated
fashion, but converges to a finite value when the Stokes time is
small, due to the balancing in the power counting in $\tau^{-1}$ and
to the finite limit of $C$ (\ref{ed:C}).

\subsection{Comparison with known results} 
\label{sec:comp}

Parallel flows are an interesting example also because it is possible to compare the perturbative 
results of section~\ref{sec:pf} to the results obtained by different non-perturbative methods in 
\cite{BMM17}.
The comparison is possible only under certain conditions. Namely we need to assume that we can analytically
continue the above expressions of the eddy diffusivity to a continuum limit
both in wave-number and frequency space. The conditions entail integration over an infinite time domain. 
We also suppose that the spectrum of the carrier velocity field is analytic in the plane defined 
by analytic continuation of the frequency variable. In such a case, if we perform the integral 
over frequencies using the Cauchy theorem the only relevant poles are those generated by the dynamics
and explicitly appearing in (\ref{pf:K0})-(\ref{pf:K2}).
It is straightforward to verify that
\begin{eqnarray}
\mathsf{K}_{\ast}^{(0)}\equiv\lim_{\mathrm{continuum}}\mathsf{K}^{(0)}
=\underline{\bm{x}}_{(1)}\otimes\underline{\bm{x}}_{(1)}\int_{\mathbb{R}^{d-1}}\!\ud\vec{m}\,|\hat{u}(\vec{m},\imath D\,m^2)|^2\;,
\nonumber
\end{eqnarray}
and consequently
\begin{eqnarray}
\lim_{D\downarrow0}\mathsf{K}_{\ast}^{(0)}=\underline{\bm{x}}_{(1)}\otimes\underline{\bm{x}}_{(1)}\int_{\mathbb{R}^{d-1}}\!\ud\vec{m}\,|\hat{u}(\vec{m},0)|^2\;.
\label{comp:K0}
\end{eqnarray}
We are thus in the position of formally recovering the first non-trivial contribution to 
the eddy diffusivity appearing on the right-hand side of (13) in \cite{BMM17}.
We emphasize the role of $D$ in the derivation of this result. In the absence of $D$, the sum 
over frequencies in (\ref{pf:K0}) contains a double pole at the origin. The sum is, however, well 
defined when $D$ is non-vanishing and yields the finite continuum limit (\ref{comp:K0}). This fact 
evinces the importance of $D$ as viscous regularization of the perturbative expansion.

We can also recover the remaining contributions to (13) in \cite{BMM17} from the continuum limit of 
(\ref{pf:K2}). Computations are conceptually straightforward but involve some algebra.
We only give the result. From the Cauchy theorem we get into
\begin{eqnarray}
\lim_{\mathrm{continuum}}\mathsf{K}^{(2)}=\mathsf{I}k+\underline{\bm{x}}_{(1)}\otimes\underline{\bm{x}}_{(1)}k\,\tau\,
\int_{\mathbb{R}^{d-1}}\!\ud\vec{m}\,m^2 \Big{(}|\hat{u}(\vec{m},\imath D\,m^2)|^2-|\hat{u}(\vec{m},\imath(D\,m^2+\tau^{-1}))|^2\Big{)}\;.
\nonumber
\end{eqnarray}
In the limit $D\downarrow0$, we recognize again after some algebra that this expression is the 
Fourier representation corresponding to the remaining terms on the right-hand side of (13) in \cite{BMM17}.
 
\subsection{Kolmogorov flows}
 
A Kolmogorov flow is a specific instance of parallel flow, with spatial sinusoidal dependence 
on one only coordinate which can be aligned with $x_d$ \cite{O83}. It reads:
 \begin{equation*}
  \bm{u}(\xt)=\underline{\bm{x}}_1U\cos(m_{\ell}x_d)\mathcal{U}(t)\;,
 \end{equation*}
with $U$ the characteristic fluid velocity scale and $m_{\ell}\equiv2\pi/\ell$. The non-dimensional function $\mathcal{U}(t)$ is 
usually taken as unity in the basic steady version, but it can also be assumed as a random function.
 
\subsubsection{Constant flow} 
\label{sec:constant}

If $\mathcal{U}(t)=1\ \forall t$, then 
 $\hat{u}(\vec{n}_{\vec{m}},n_{\omega})=U\delta_{n_{\omega},0}\delta_{n_{m_2},0}\ldots\delta_{n_{m_{d-1}},0}(\delta_{n_{m_d},1}+\delta_{n_{m_d},-1})/2$.
 By denoting with $\vartheta$ the angle between $\bm{g}$ and $\underline{\bm{x}}_d$, we get:
 \begin{equation} \label{K0K}
  K_{ij}^{(0)}=\delta_{i1}\delta_{j1}\frac{U^2}{2D\,m_{\ell}^2}\;,
 \end{equation}
 \begin{equation} \label{K2K}
  \alpha K_{ij}^{(2)}=\delta_{i j}\kappa-\delta_{i1}\delta_{j1}\frac{U^2}{2D^2m_{\ell}^2(D\,m_{\ell}^2\tau+1)}\kappa\;,
 \end{equation}
 \begin{equation} \label{K4K}
  \alpha^2K_{ij}^{(4)}=\delta_{i1}\delta_{j1}\left[\frac{U^2\tau^2g^2(\cos\vartheta)^2}{2D^3m_{\ell}^4}(1-\beta)^2+\frac{U^2(3D\,m_{\ell}^2\tau+2)}{2D^3m_{\ell}^2(D\,m_{\ell}^2\tau+1)^2(D\,m_{\ell}^2\tau+2)}\kappa^2\right]\;.
 \end{equation}
It is interesting to notice that only the component of diffusivity parallel to the flow is modified, 
but that such a modification depends on the angle between the vertical and the $x_d$ directions 
(and not on the one between gravity and $x_1$). 
Equation (\ref{K0K}) tells us that, for neutrally-buoyant particles, the parallel effective diffusivity 
is inversely proportional to the Brownian one. 

The leading-order correction (\ref{K2K}) is linear in the Brownian diffusivity $\kappa$, and consists of an 
isotropic constant component minus a parallel one which is a decreasing function of both $\tau$ and
$D$. The overall sign of (\ref{K2K}) is not defined a priori:
 \begin{equation*}
  \alpha K_{11}^{(2)}\ge0\ \Longleftrightarrow\ \tau\ge\frac{1}{D\,m_{\ell}^2}\left(\frac{U^2}{2D^2m_{\ell}^2}-1\right)\;.
 \end{equation*}
 
The main sub-leading correction from (\ref{K4K}) 
is always positive in 
the parallel direction, second-order in either $1-\beta$ or $\kappa$, a growing function of gravity, maximum for $\vartheta=0$ or $=\pi$ 
(a horizontal flow) and minimum for $\vartheta=\pi/2$ (surely corresponding to a vertical flow only in $d=2$,
but not necessarily in $d=3$). 
Numerically, one also infers the presence of a minimum in $\tau$ and a monotonic dampening in $D$.

\begin{figure}
  \includegraphics{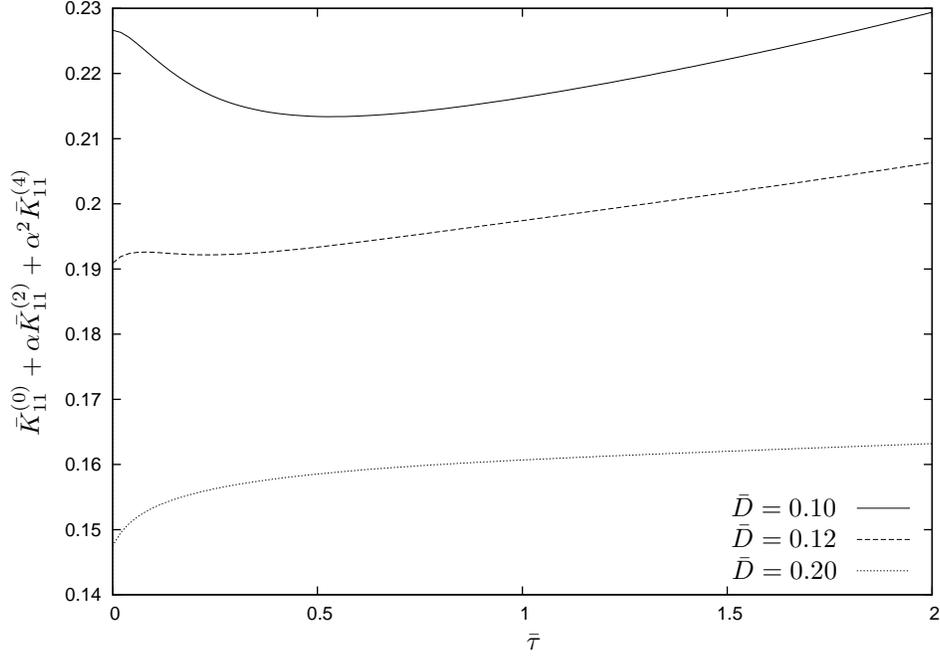}
  \caption{Parallel component (for the constant Kolmogorov flow) of the rescaled effective diffusivity up to working order,
   as a function of the rescaled Stokes time, plotted for different values of the rescaled Brownian diffusivity.}
  \label{fig}
 \end{figure}

Figure~\ref{fig} shows the parallel component of the sum of expressions (\ref{K0K})+(\ref{K2K})+(\ref{K4K})
as a function of the Stokes time. Plotted are the situations corresponding to three values of the Brownian diffusivity,
all small in accordance to its character of regularization parameter.
The quantities are drawn in units of $\ell$ and $U$:
$\bar{\tau}\equiv\tau/(\ell/U)$ and $\bar{D}\equiv D/(\ell U)$ (and similarly for $\bar{\mathsf{K}}$).
It is interesting to notice how modifications
of $D$ bring about radical changes in the concavity and initial slope of the curves.
One of the relevant results of the present work is that it allows for the investigation
at finite values of $\tau$, without any perturbative expansion in such a parameter.
As a consistency check, we verified (not shown here) that both ratios
$(\alpha^2K_{11}^{(4)})/(\alpha K_{11}^{(2)})$ and $(\alpha K_{11}^{(2)})/K_{11}^{(0)}$ are small in this situation.
%%%%%%%%%%
%%%%%%%%%%

\paragraph{Illustrative example} To shed more light on (\ref{K0K}),
we contrast it with the solution of (\ref{basic}) in the limit case $\beta=1$ and $\kappa=0$. 
In this case $\dot{\bm{\mathcal{V}}}(t)=-\bm{\mathcal{V}}(t)/\tau$ so that the initial velocity of the particle becomes irrelevant for the computation. 
Assuming moreover $D=0$, the inertial-particle dynamics reduces to that of a tracer in the absence of noise:
 \begin{equation} \label{kolcon}
  \dot{\bm{\mathcal{X}}}(t)=\bm{u}(\bm{\mathcal{X}}(t),t)=\underline{\bm{x}}_1U\cos(m_{\ell}\mathcal{X}_d(t))\;.
 \end{equation}
We can explicitly integrate (\ref{kolcon}). For $\mathcal{X}_d(t)=x_{d}$ 
and $\mathcal{X}_1(0)=0$, we get into
\begin{eqnarray}
\mathcal{X}_1(t)=U\cos(m_{\ell}\,x_d)\,t\;.
\nonumber
\end{eqnarray} 
The transport properties are then defined by averages on the initial positions of the particles. 
In particular, averaging over the $x_d$ coordinate gives a zero terminal velocity, as expected.
The effective diffusivity blows up:
 \begin{equation*}
  K_{11}=\lim_{t\to+\infty}\frac{\langle\mathcal{X}_1^2(t)\rangle-\langle\mathcal{X}_1(t)\rangle^2}{2t}\propto\lim_{t\to+\infty}\frac{t^2}{t}=+\infty\;.
 \end{equation*}
This infinite result is in perfect agreement with the limit $D\to0$ in (\ref{K0K}), and shows that
deterministic flows evolving random initial data can give rise to ballistic diffusion.

\subsubsection{Random flow} 
\label{sec:random}

We now suppose that $\mathcal{U}(t)$ is a random, stationary Gaussian process with zero mean, and correlation 
function \cite{MY75}
 \begin{equation*}
  \mathcal{W}(t)\equiv\langle\mathcal{U}(t)\mathcal{U}(0)\rangle=\exp(-\Gamma|t|)\cos(\gamma t)\;.
 \end{equation*}
The rates $\Gamma$ and $\gamma$ respectively measure the inverse vortex life time and the re-circulation degree. In other words, 
$\Gamma$ and $\gamma$ are inversely and directly proportional to the Kubo and Strouhal numbers, respectively.
In the spirit of a comparison with the time-periodic case, we may identify $\gamma$ with $2\pi/\mathcal{T}$.
We may, however, consider the case of an infinite temporal domain. We interpret in such a case the rate $\Gamma$ as a 
regularization dampening correlations. 
We recover the constant-Kolmogorov case of subsection~\ref{sec:constant} by letting $\Gamma\to0\gets\gamma$ and paying attention 
to the order of the limits in the non-diffusive case $D\to0$.
We apply the results of the previous sections, by replacing the temporal Fourier series with integrals,
and also substituting time averages over one period with statistical averages. With these provisos, we replace
 \begin{equation*}
  \int_0^{\mathcal{T}}\!\frac{\ud t}{\mathcal{T}}\sum_{n_{\omega},n_{\omega'}\in\mathbb{Z}}\!\hat{u}(\vec{n}_{\vec{m}},n_{\omega})
\hat{u}(\vec{n}_{\vec{m}'},n_{\omega'})\exp[\imath(\omega+\omega')t]\mapsto
\int_{\mathbb{R}}\!\ud\omega\,|\hat{u}_{\star}(\vec{n}_{\vec{m}},\omega)|^2\;,
 \end{equation*}
 with $|\hat{u}_{\star}(\vec{n}_{\vec{m}},\omega)|^2=U^2\hat{\mathcal{W}}(\omega)\delta_{n_{m_2},0}\ldots\delta_{n_{m_{d-1}},0}(\delta_{n_{m_d},1}+\delta_{n_{m_d},-1})/4$ and
 \begin{equation*}
  \hat{\mathcal{W}}(\omega)=\frac{\Gamma}{\sqrt{2\pi}}\left[\frac{1}{\Gamma^2+(\omega+\gamma)^2}+\frac{1}{\Gamma^2+(\omega-\gamma)^2}\right]\;.
 \end{equation*}
 Consequently,
 \begin{equation} \label{cappa0K}
  K_{i j}^{(0)}=\delta_{i1}\,\delta_{j1}\frac{U^2\,\Omega}{2\,(\Omega^2+\gamma^2)}\;,
 \end{equation}
 \begin{equation} \label{cappa2K}
  \alpha K_{i j}^{(2)}=
\delta_{i j}\kappa-\delta_{i1}\delta_{j1}
\frac{U^2\,m_{\ell}^2}{2\,\tau^2}\frac{\Omega\,(\Omega^2-3\,\gamma^2)\,\tau+(\Omega^2-\gamma^2)}{(\Omega^2+\gamma^2)^2[(\Omega+\tau^{-1})^2+\gamma^2]}\kappa\;,
 \end{equation}
 \begin{equation} 
\label{cappa4K}
  \alpha^2 K_{i j}^{(4)}=\delta_{i 1}\,\delta_{j 1}
\left[U^2\,\tau^2\,g^2\,m_{\ell}^2(\cos\vartheta)^2\frac{\Omega\,(\Omega^2-3\,\gamma^2)}{2(\Omega^2+\gamma^2)^3}(1-\beta)^2+\mathcal{C}\,\kappa^2\right]\;.
 \end{equation}
In the above expressions we defined $\Omega$ as the weighted sum of the Brownian diffusivity and of the inverse 
vortex life time,
\begin{eqnarray}
\Omega\equiv D\,m_{\ell}^2+\Gamma\;,
\nonumber
\end{eqnarray}
 and
 \begin{eqnarray*}
  \mathcal{C}&=&\frac{U^2m_{\ell}^4}{2\tau^2(\Omega^2+\gamma^2)^3[(\Omega+\tau^{-1})^2+\gamma^2]^2[(\Omega+2\tau^{-1})^2+\gamma^2]}\times\\
  &&\times[3\Omega(\Omega^6-9\Omega^4\gamma^2-5\Omega^2\gamma^4+5\gamma^6)+2(7\Omega^6-45\Omega^4\gamma^2-15\Omega^2\gamma^4+5\gamma^6)\tau^{-1}+\Omega\times\\
  &&\quad\times(23\Omega^4-110\Omega^2\gamma^2-5\gamma^4)\tau^{-2}+4(4\Omega^4-15\Omega^2\gamma^2+\gamma^4)\tau^{-3}+4\Omega(\Omega^2-3\gamma^2)\tau^{-4}]\;.
 \end{eqnarray*}
Equation (\ref{cappa0K}) tells us that, for neutrally-buoyant particles, the parallel effective diffusivity 
is a decreasing function of the re-circulation degree. Furthermore the effective diffusivity vanishes for very 
small or very large $\Omega$, and is maximum for $\Omega=\gamma$. 

The leading correction (\ref{cappa2K}) is again linear in the small parameter $\kappa$. It consists of an isotropic constant 
component and a parallel one which is a complicated function of $\tau$, $\Omega$ and $\gamma$.
Its overall sign is not defined a priori.

The main sub-leading correction from (\ref{cappa4K})
can have either sign in the parallel direction, 
is second-order in both $1-\beta$ and $\kappa$, and is a growing function of gravity, maximum for $\vartheta=0$ or $=\pi$ and minimum 
for $\vartheta=\pi/2$.

\paragraph{Illustrative example} Finally, it is again expedient to contrast (\ref{cappa0K}) with the solution of (\ref{basic}) 
in the limit case $\beta=1$ and $\kappa=0$. 
Neglecting as in subsection~\ref{sec:constant} the initial velocity and imposing $D=0$, the inertial 
particle dynamics reduces for $\mathcal{X}_{d}(0)=x_{d}$ to:
\begin{eqnarray}
\dot{\bm{\mathcal{X}}}(t)=\underline{\bm{x}}_1U\cos(m_{\ell}\,x_d)\,\mathcal{U}(t)\;.
\nonumber
\end{eqnarray}
Let us then focus on the following two cases.
\begin{enumerate}
\item Case $\gamma\neq0=\Gamma$. This is a time-oscillating flow with a random phase $\varpi$ distributed 
uniformly on $[0,2\pi)$:
 \begin{equation} 
\label{kolsin}
  \mathcal{U}(t)=\sqrt{2}\cos(\gamma t+\varpi)\ \Longrightarrow\ \langle\mathcal{U}(t)\rangle=0
\quad\&\quad\langle\mathcal{U}(t)\mathcal{U}(t')\rangle=\cos[\gamma(t-t')]\;.
 \end{equation}
For $\mathcal{X}_{1}(0)$, the solution of (\ref{kolsin}) is
\begin{eqnarray}
\mathcal{X}_1(t)=\sqrt{2}U\cos(m_{\ell}\,x_d)\sin(\gamma t+\varpi)/\gamma
\nonumber
\end{eqnarray}
whence we find a vanishing terminal velocity, and for the effective diffusivity
 \begin{equation*}
  K_{11}=\lim_{t\to+\infty}\frac{\langle\mathcal{X}_1^2(t)\rangle-\langle\mathcal{X}_1(t)\rangle^2}{2t}\propto\lim_{t\to+\infty}\frac{\mathrm{const.}}{t}=0\;.
 \end{equation*}
This zero result is in perfect agreement with the situation $D=0=\Gamma$ in (\ref{cappa0K}) (i.e.\ $\Omega=0$) for finite $\gamma$.
\item Case $\Gamma\neq0=\gamma$. This corresponds to a telegraph process, with random initial condition $\mathcal{U}(0)$ and only 
two possible states $\pm1$, with the jump distribution described by a Poisson process $\varpi(t)$ with parameter $\Gamma/2$:
 \begin{eqnarray} 
\label{kolexp}
  &\displaystyle\mathcal{U}(t)=(-1)^{(\mathcal{X}(0)-1)/2+\varpi(t)}
\quad\&\quad
\mathrm{prob}[\varpi(t)=\mathcal{I}]=\exp(-\Gamma t/2)\frac{(\Gamma t/2)^{\mathcal{I}}}{\mathcal{I}!}
\nonumber\\
  &\Longrightarrow\ \langle\mathcal{U}(t)\rangle=0\quad\&\quad\langle\mathcal{U}(t)\mathcal{U}(t')\rangle=\exp(-\Gamma|t-t'|)\;.
 \end{eqnarray}
We find again as expected a zero terminal velocity. We can also use (\ref{kolexp}) to derive the 
effective diffusivity:
 \begin{equation*}
  K_{11}=\lim_{t\to+\infty}\int_0^t\!\ud t'\langle u_1(\bm{\mathcal{X}}(t),t)u_1(\bm{\mathcal{X}}(t'),t')\rangle\propto\lim_{t\to+\infty}\frac{1-\exp(-\Gamma t)}{\Gamma}=\Gamma^{-1}\;.
 \end{equation*}
This result is in perfect agreement with the situation $D=0=\gamma$ in (\ref{cappa0K}) with $\Omega=\Gamma$, and 
consistently implies an infinite diffusivity in the limit of vanishing $\Gamma$.
\end{enumerate}

\section{Conclusions} 
\label{sec:cp}

Multi-scale perturbation theory confers a precise mathematical meaning to
the distinction between small- and large-scale hydrodynamic degrees of freedom. As a consequence
experimentally relevant indicators such as terminal velocity and eddy diffusivity can be
expressed as well-defined averages over microscopic degrees of freedom. The explicit evaluation
of these quantities is of crucial importance for applications e.g.\ in micro-meteorology
\cite{GRSLAVM16}. The evaluation requires, however, to resolve the small-scale
dynamics. This remains a task computationally very challenging in spite of the
dimensional reduction operated by multi-scale methods.

The perturbative scheme of the present contribution further simplifies this task.
It differs from the over-damped expansion applied with the same purpose in \cite{MAMM12}. Using the
deviation from unity of the added-mass ratio as expansion parameter allows us to resolve
small-scale co-velocity and position dynamics on the same footing. The consequent and notable difference with
the over-damped expansion, is that we can handle the analysis of small-scale degrees of freedom
by means of a regular perturbation theory rather than by a secondary multi-scale expansion.
This is an advantage inasmuch multi-scale expansions, similarly to the classical Chapman--Enskog procedure,
encounter in general obstructions at orders higher than the second. The reason relies, roughly speaking, in the
generic occurrence of short-wave instabilities. In the framework of multi-scale expansions these instabilities
appear in connection with the appearance of secular terms (see \S~10.4.1 of \cite{R89} for an explicit illustration
and \cite{B06} for a broad discussion).

Here, we compute higher-order corrections by solving well-posed equations without the need of
partial resummations to cancel the occurrence of secular terms.
As a consequence the expansion appears, at least formally, free of short-wave instabilities.
In this sense, and in the context of inertial particles with added mass, it may be regarded as a way
to circumvent short-wave instabilities affecting hydrodynamic perturbative expansions in more general contexts.

Finally, it is also possible to envisage extensions of the present methods such as considering compressible carrier
flows or, a considerably more challenging task, applying it to general inertial models including Basset-type
history terms, as well as other corrections due to the contributions by Fax\'en, Oseen and Saffman.

 \begin{acknowledgments}
PMG acknowledges support by Academy of
Finland via the Centre of Excellence in Analysis and Dynamics Research
(project No.\ 271983), and the \textbf{AtMath Collaboration} at the
University of Helsinki. MMA and SMAG were partially supported by CMUP
(UID/MAT/00144/2013), which is funded by FCT (Portugal) with national (MEC) and
European structural funds (FEDER), under the partnership agreement PT2020; and also
by Project STRIDE - NORTE-01-0145-FEDER-000033, funded by ERDF NORTE 2020.
This article is based upon work from COST Action MP1305, supported by COST (European Cooperation in Science and Technology).
 \end{acknowledgments}

 \appendix

\section{Hermite polynomials} 
\label{ap:Hermite}
 
We write the expression in Cartesian components of the Hermite polynomials used in the text:
 \begin{eqnarray*}
  &H^{(2)}_{ij}(\bm{y})\equiv y_i\,y_j-k\tau^{-1}\delta_{ij}\;,
  \qquad H^{(3)}_{ijl}(\bm{y})\equiv y_i\,y_j\,y_l
-k\tau^{-1}\displaystyle{\sum_{\sigma}}y_{\sigma(i)}\delta_{\sigma(j)\sigma(l)}\;,\\
  &H^{(4)}_{ijlh}(\bm{y})\equiv y_i\,y_j\,y_l\,y_h
-k\,\tau^{-1}\displaystyle{\sum_{\sigma}}y_{\sigma(i)}y_{\sigma(j)}\delta_{\sigma(l)\sigma(h)}
+k^2\tau^{-2}\displaystyle{\sum_{\sigma}}\delta_{\sigma(i)\sigma(j)}\delta_{\sigma(l)\sigma(h)}\;.
 \end{eqnarray*}
As in the main text the sum over $\sigma$ ranges over index permutations without ordering. Hermite
polynomials are orthogonal with respect to the scalar product defined by the integral with respect to
the Gaussian measure (\ref{pt:Gaussian}) --- see \cite{G49} for details.

\section{Floquet theory} 
\label{ap:Floquet}
 
We expect that the semigroup
$\mathcal{G}_{t}$ solution of 
\begin{subequations}
\begin{eqnarray*}
&&\breve{\mathcal{M}}_{\xt}^{\dag}\mathcal{G}_{t}(\bm{x},\bm{x}_{\bullet})=0\;,
\\
&&\lim_{t\downarrow 0}\mathcal{G}_{t}(\bm{x},\bm{x}_{\bullet})=\delta_{\mathbb{B}}(\bm{x}-\bm{x}_{\bullet})
\end{eqnarray*}
\end{subequations}  
(where $\delta_{\mathbb{B}}$ is the Dirac-$\delta$ on the torus $\mathbb{T}^{d}(\mathbb{B})$), admits the factorization
\begin{equation*}
\mathcal{G}_{t}(\bm{x},\bm{x}_{\bullet})=\mathcal{P}_{t}(\bm{x},\bm{x}_{\bullet})
\,\exp(-\mathcal{F}\,t)(\bm{x},\bm{x}_{\bullet})\;.
\end{equation*}
The semigroup $\mathcal{P}_{t}$ satisfies $\mathcal{P}_{t+\mathcal{T}}=\mathcal{P}_{t}$ for all $t$. The operator $\mathcal{F}$ is time
autonomous. 
By the ergodicity assumption $\exp(-\mathcal{F}\,t)$ is contractive on $\mathbb{L}^{2}(\mathbb{T}^{d})$
except when acting on constant functions, i.e.\ on the elements of the kernel of 
$\breve{\mathcal{M}}_{\xt}$ and $\breve{\mathcal{M}}^{\dag}_{\xt}$. These considerations lead us to expect 
that the spectrum ($\operatorname{Sp}$) of $\breve{\mathcal{M}}_{\xt}^{\dag}$
on $\mathbb{L}^{2}(\mathbb{T}^{d+1})$ satisfies
\begin{eqnarray}
\mathrm{Sp}\,\breve{\mathcal{M}}^{\dag}_{\xt}
=\operatorname{Sp}\mathcal{F}\pm\frac{2\,\pi\,\imath\,n}{\mathcal{T}}\qquad\forall\,n\in\,\mathbb{N}\;.
\nonumber
\end{eqnarray}

\section{Explicit calculations for parallel flows} 
\label{ap:parallel}

\subsection{Steady state}

It is readily seen that the only non vanishing component of (\ref{ss:P1}) 
becomes in Fourier space:
\begin{eqnarray}
  \hat{P}_{1}^{(1:1)}(\nmo)=-(\imath\,\omega+D\,m^2)^{-1}J\,\tau^{-1}\,\hat{u}_{1}(\nmo)\;.
\nonumber
\end{eqnarray}
As $P^{(1:1)}$ is divergenceless, we immediately find that:
\begin{eqnarray}
P^{(2:0)}=\bm{W}^{(2)}=0\;.
\nonumber
\end{eqnarray}
Inspection of (\ref{ss:P22}) and (\ref{ss:P31}) shows that
\begin{eqnarray}
\partial_{x_{i}}\partial_{x_{j}}P^{(2:2)}_{i j}=\partial_{x_{i}}P^{(3:1)}_{i}=0\;,
\nonumber
\end{eqnarray}
and that only $P^{(3:1)}_{1}$ and $P^{(2:2)}_{1 1}$ are non-vanishing.
Using this result, we conclude:
\begin{eqnarray}
P^{(4:0)}=W^{(4)}_{i}=0\;.
\nonumber
\end{eqnarray}

\subsection{Auxiliary vector}

Once we take into account initial and solvability conditions, (\ref{ed:zero}) 
reduces to a diffusion equation for the first component $Q^{(0:0)}_{1}$ as a function of $\vec{x}=(x_{2},\dots,x_{d})$. 
In Fourier space this observation transduces into 
 \begin{equation*}
  \hat{Q}_{1}^{(0:0)}(\nmo)=-(\imath\,\omega+D\,m^2)^{-1}\hat{u}_{1}(\nmo)\;,
 \end{equation*}
whence we readily obtain (\ref{pf:K0}).

In order to compute the leading-order correction to the eddy diffusivity we need to compute 
the vector $\mathcal{Q}_{i}^{(1:1)}\equiv\partial_{x_{j}}Q^{(1,1)}_{ij}$.
Since $P^{(3:1)}_{i}\partial_{x_{i}}u_{1}=P^{(3:1)}_{1}\partial_{x_{1}}u_{1}=0$, only $\mathcal{Q}^{(1:1)}_{1}$ is non-vanishing 
and obeys
\begin{eqnarray}
(\partial_t-D\,\partial_{\bm{x}}^{2}+\tau^{-1})\,\mathcal{Q}^{(1:1)}_{1}(\vxt)=-\,\partial_{\vec{x}}^{2}\,Q^{(0:0)}_{1}(\vxt)\;.
\nonumber
\end{eqnarray}
Turning to Fourier space we find:
 \begin{eqnarray} 
\label{parallel:h11}
  \hat{\mathcal{Q}}_{1}^{(1:1)}(\nmo)=(\imath\,\omega+D\,m^2+\tau^{-1})^{-1}\,m^2\,\hat{Q}^{(0:0)}_{1}(\nmo)\;.
 \end{eqnarray}
Upon inserting in (\ref{ed:Q20tilde}), we find that the only non-vanishing contribution comes from
 \begin{eqnarray} 
\label{parallel:Q20}
  \hat{R}^{(2:0)}_{1}(\nmo)=
-\,(\imath\,\omega+D\,m^2)^{-1}\Big{(}\imath\,J\,\vec{g}\cdot\vec{m}\,\hat{Q}^{(0:0)}_{1}(\nmo)+k\,\tau^{-1}\,\hat{\mathcal{Q}}_{1}^{(1:1)}(\nmo)\Big{)}\;.
 \end{eqnarray}
Using (\ref{parallel:h11}), (\ref{parallel:Q20}) we obtain 
 \begin{equation*}
\mathsf{K}^{(2)}=k\,\mathsf{I}
-\underline{\bm{x}}_1\otimes\underline{\bm{x}}_1\sum_{\vec{n}_{\vec{m}},n_{\omega}}
\frac{k\,\tau^{-1}\,m^2+\imath J\tau\vec{g}\cdot\vec{m}(\imath\omega+D\,m^2+\tau^{-1})}
{(\imath\omega+D\,m^2)^2(\imath\omega+D\,m^2+\tau^{-1})}|\hat{u}(\nmo)|^2\;.
\end{equation*}  
From this expression we obtain (\ref{pf:K2}) upon exploiting the fact 
that $\hat{u}(\nmo)$ is the Fourier transform of a real-valued vector field.

In order to compute the second-order correction we need $\mathcal{Q}_{i}^{(3:1)}(\xt)\equiv \partial_{x_{j}}Q^{(3:1)}_{j i}(\xt)$,
whose only non-vanishing component is determined by the solution of
\begin{equation*} 
  (\partial_t-D\,\partial_{\bm{x}}^{2}+\tau^{-1})\,\mathcal{Q}_{1}^{(3:1)}=
-\partial_{\vec{x}}^2Q^{(2:0)}_{1}
-J\,k^{-1}\bm{g}\cdot\partial_{\bm{x}}\mathcal{Q}^{(1:1)}_{1}
-2\,k\,\tau^{-1}\,\partial_{x_{i}}\partial_{x_{j}}Q^{(2:2)}_{ij1}\;,
 \end{equation*}
and
 \begin{equation*}
  (\partial_t-D\,\partial_{\bm{x}}^{2}+2\tau^{-1})\partial_{x_{i}}\partial_{x_{j}}\,Q^{(2:2)}_{ij1}=-\partial_{\vec{x}}^{2}\mathcal{Q}^{(1:1)}_{1}\;.
 \end{equation*}
Equipped with the solutions of the foregoing equations, we can explicitly
determine
 \begin{eqnarray*}
  \hat{Q}_{j}^{(4:0)}(\nmo)=-(\imath\omega+D\,m^2)^{-1}k\,\tau^{-1}\,\hat{\mathcal{Q}}_{j}^{(3:1)}(\nmo)\;,
 \end{eqnarray*} 
and therefore
 \begin{eqnarray*}
  \mathsf{K}^{(4)}&=&-\sum_{\vec{n}_{\vec{m}},n_{\omega}}\hat{\bm{u}}(-\vec{n}_{\vec{m}},-n_{\omega})\otimes \hat{\bm{Q}}^{(4:0)}(\nmo)\\
  &=&-\underline{\bm{x}}_1\otimes\underline{\bm{x}}_1\,k\,\tau^{-1}\sum_{\vec{n}_{\vec{m}},n_{\omega}}(\imath\,\omega+D\,m^2)^{-1}\hat{u}(-\vec{n}_{\vec{m}},-n_{\omega})
\,\hat{\mathcal{Q}}^{(3:1)}_{1}(\nmo)\;.
 \end{eqnarray*}
From this expression we get into (\ref{pf:K4}) after some straightforward 
yet lengthy algebra.

%merlin.mbs apsrev4-1.bst 2010-07-25 4.21a (PWD, AO, DPC) hacked
%Control: key (0)
%Control: author (0) dotless jnrlst
%Control: editor formatted (1) identically to author
%Control: production of article title (0) allowed
%Control: page (1) range
%Control: year (0) verbatim
%Control: production of eprint (0) enabled
%

% \bibliography{bkPRFbib}

\end{document}